\documentclass[prd,12pt]{revtex4}
\pdfoutput=1 
\usepackage{amsmath}
\usepackage{graphicx}
\usepackage{subfigure}
\usepackage{bm}
\usepackage{slashed}

\def\d{\mathrm{d}}
\def\beq{\begin{equation}}
\def\eeq{\end{equation}}
\def\bea{\begin{eqnarray}}
\def\eea{\end{eqnarray}}

\def\nnb{\nonumber}
\def\ga{\left(}
\def\dr{\right)}

\def\nnb{\nonumber}

\def\ba{\begin{array}}
\def\ea{\end{array}}

\begin{document}
\title{Predicting the $B \to K^*$ form factors in light-cone QCD}

\author{M. Ahmady}
\affiliation{Department of Physics, Mount Allison University, Sackville, N-B. E46 1E6, Canada}
\email{mahmady@mta.ca,recampbell@mta.ca} 

\author{R. Campbell}
\affiliation{Department of Physics, Mount Allison University, Sackville, N-B. E46 1E6, Canada}
\email{recampbell@mta.ca} 

\author{S. Lord}
\affiliation{D\'epartement de Math\'ematiques et Statistique, Universit\'e de Moncton,
Moncton, N-B. E1A 3E9, Canada}
\email{esl8420@umoncton.ca}

\author{R. Sandapen}
\affiliation{D\'epartement de Physique et d'Astronomie, Universit\'e de Moncton,
Moncton, N-B. E1A 3E9, Canada
\& \\
Department of Physics, Mount Allison University, Sackville, N-B. E46 1E6, Canada }
\email{ruben.sandapen@umoncton.ca} 

\begin{abstract}
 We use QCD light-cone sum rules  with holographic anti de Sitter/Chromodynamics (AdS/QCD)  Distribution Amplitudes (DAs)  for the $K^*$ meson in order to predict the full set of seven $B \to K^*$ transition form factors for  intermediate-to-high recoil of the vector meson. We provide simple parametrizations for the form factors that fit our AdS/QCD predictions. We also provide parametrizations that fit both our AdS/QCD predictions and the most recent lattice data for low recoil.   We use our form factors to predict the differential and total branching fraction of the rare dileptonic decay $B \to K^* \mu^+ \mu^-$ which we compare to the recent LHCb data.

\end{abstract}


\maketitle

\section{Introduction}

$B$ meson decays to $K^{(*)}$ are mediated by the penguin-induced flavor changing neutral current $b \to s$ transition which is an excellent venue for precision tests of the Standard Model (SM) and for probing new physics beyond the SM. 
Moreover, three-body decays like $B \to K^{*} (\to K \pi) l^+ l^-$ provide a plethora of angular observables, some of which  may be sensitive to new physics. These considerations, along with the fact that the LHCb detector has a high efficiency in detecting muons, explains the fact that the decay $B \to K^{*} \mu^+ \mu^-$ is presently generating a great deal of 
experimental \cite{Aaltonen:2011ja,Lees:2012tva,Wei:2009zv,Aaij:2013qta,Aaij:2012cq,Aaij:2013iag,LHCb-CONF-2012-008} and theoretical \cite{Altmannshofer:2013foa,Descotes-Genon:2013wba,Hurth:2013ssa,Descotes-Genon:2013vna,Gauld:2013qba,Buras:2013qja,Gauld:2013qja} activity.

In a previous paper \cite{Ahmady:2013cva}, we  derived four holographic AdS/QCD DAs for the $K^*$ meson: two twist-$2$ DAs, one for each polarization of the $K^*$; and two twist-$3$ DAs, vector and axial vector, for the transversely polarized $K^*$. We 
used the transverse DAs in order to predict the branching ratio and the power-suppressed isospin asymmetry of the radiative decay $B \to K^* \gamma$. We found that the transverse twist-$2$ AdS/QCD DA offers an advantage in that it avoids the end-point divergence encountered when computing the isospin asymmetry using the transverse twist-$2$ sum rules DA. Our predictions were consistent with experiment \cite{Ahmady:2013cva}.  

Our goal in this paper is to use both the longitudinal and transverse twist-$2$ AdS/QCD  DAs in QCD light-cone sum rules \cite{Ali:1993vd,Aliev:1996hb,Ball:2004rg} in order to compute the seven $B \to K^*$ transition form factors\footnote{A recent computation of the $B \to \rho$ form factors can be found in Ref. \cite{Ahmady:2013cga}.}. We shall then use these form factors to predict the differential branching ratio of the dileptonic decay $B \to K^* \mu^+ \mu^-$ which we shall compare with the recent data released by the LHCb collaboration \cite{Aaij:2012cq,LHCb-CONF-2012-008}. The angular analysis of this decay is particularly interesting since the LHCb collaboration recently reported \cite{Aaij:2013qta} a $3.7\sigma$ discrepancy between one angular observable at high recoil and the Standard Model prediction  of Ref. \cite{Descotes-Genon:2013vna} .  Unlike the differential branching fraction, this particular angular observable is largely free from the hadronic uncertainties related to the form factors. It is not yet clear whether this discrepancy is caused by new physics or is due  to statistical fluctuations and/or other theoretical uncertainties  \cite{Hurth:2013ssa}. If the LHCb anomaly is due to new physics phenomena, they could perhaps be revealed in other observables including the differential and total branching fraction, provided we have a good understanding of the uncertainties in the $B \to K^*$ form factors.

\section{Form factors}
The seven $B \to K^*$ transition form factors $V,A_0,A_1,A_2,T_1,T_2$ and $T_3$ are defined by the following expressions \cite{Horgan:2013hoa}:
\bea
\langle K^* (k,\varepsilon)|\bar{s} \gamma^\mu(1-\gamma^5 )b | B(p) \rangle &=& \frac{2i V(q^2)}{m_B + m_{K^*}} \epsilon^{\mu \nu \rho \sigma} \varepsilon^*_{\nu} k_{\rho} p_{\sigma} -2m_{K^*} A_0(q^2) \frac{\varepsilon^* \cdot q}{q^2} q^{\mu}  \nonumber \\
&-& (m_B + m_{K^*}) A_1(q^2) \left(\varepsilon^{\mu *}- \frac{\varepsilon^* \cdot q q^{\mu}}{q^2} \right) \nonumber \\
&+& A_2(q^2) \frac{\varepsilon^* \cdot q}{m_B + m_{K^*}}  \left[ (p+k)^{\mu} - \frac{m_B^2 - m_{K^*}}{q^2} q^{\mu} \right] 
\eea
and
\bea
q_{\nu} \langle K^* (k,\varepsilon)|\bar{s} \sigma^{\mu \nu} (1-\gamma^5 )b | B(p) \rangle &=& 2 T_1(q^2) \epsilon^{\mu \nu \rho \sigma} \varepsilon^*_{\nu} p_{\rho} k_{\sigma} \nonumber \\
&-& i T_2(q^2)[(\varepsilon^* \cdot q)(p+k)_{\mu}-\varepsilon_{\mu}^*(m_B^2-m_{K^*}^2)] \nonumber \\
&-& iT_3(q^2) (\varepsilon^* \cdot q) \left[ \frac{q^2}{m_B^2-m_{K^*}^2} (p+k)_{\mu} -q_{\mu}  \right] 
\eea
where $q=p-k$ is the $4$-momentum transfer  and $\varepsilon$ is the polarization $4$-vector of the $K^*$. 

At low-to-intermediate values of $q^2$, these form factors can be computed using QCD light-cone sum rules \cite{Ali:1993vd,Aliev:1996hb,Ball:2004rg} . Here we shall use the light-cone sum rules derived in Ref. \cite{Aliev:1996hb}:
\bea
V(q^2) &=& \left(\frac{m_B + m_{K^*}}{2}\right) \left(\frac{m_b}{f_B m_B^2} \right)
\, \exp \left(\frac{m_B^2}{M^2}\right)
\int_\delta^1 \d u \, \exp \ga - \frac{m_b^2 + p^2 u
\bar u - q^2 \bar u}{u M^2} \dr \nnb \\ 
&& \times \left\{ m_b  m_{K^*} \frac{f_{K^*}g_\perp^{(a)}(u)}{2 u^2 M^2} +
\frac{f_{K^*}^\perp \phi_\perp(u)}{u} \right\}~, 
\label{LCSRV}
\eea
\bea
A_1(q^2) &=& \left(\frac{1}{m_B + m_{K^*}}\right) \left( \frac{m_b}{f_B m_B^2} \right)
\, \exp \left(\frac{m_B^2}{M^2}\right)
\int_\delta^1 \d u \, \exp \ga - \frac{m_b^2 + p^2 u
\bar u - q^2 \bar u}{u M^2} \dr \nnb \\
&& \times \left\{ m_b m_{K^*} \frac{ f_{K^*}g_\perp^{(v)}(u)}{u} +
\frac{(m_b^2 - q^2 +p^2 u^2) f_{K^*}^\perp \phi_\perp(u) }{2 u^2}  
\right\}~,
\label{LCSRA1}
\eea
\bea
A_2(q^2) &=& - \, \ga m_B + m_{K^*} \dr \left(\frac{m_b}{f_B m_B^2} \right)
\, \exp{\left(\frac{m_B^2}{M^2}\right)}
\int_\delta^1 \d u \, \exp \ga - \frac{m_b^2 + p^2 u
\bar u - q^2 \bar u}{u M^2} \dr  \nnb \\
&& \times \left\{ -\frac{m_b  m_{K^*}}{u^2 M^2}
f_{K^*}\Phi_\parallel (u)   - \frac{1}{2} \, 
\frac{f_{K^*}^\perp \phi_\perp (u)}{u} \right\}~,
\label{LCSRA2}
\eea
\bea
T_1(q^2) &=& \frac{1}{4} \left( \frac{m_b}{f_B m_B^2}\right) \exp{\left(\frac{m_B^2}{M^2}\right)}
\int_\delta^1 \frac{\d u}{u}\, \exp \ga - \frac{m_b^2 + p^2 u \bar u - q^2 \bar u}{u
M^2} \dr \Bigg\{ m_b f_{K^*}^\perp \phi_\perp (u) + \nnb \\
&&\; f_{K^*} m_{K^*} \Bigg[ \Phi_\parallel (u)  +
u g_\perp^{(v)} (u)  +\frac{g_\perp^{(a)}(u)}{4} + \frac{(m_b^2 + q^2 -
p^2 u^2)g_\perp^a (u)}{4 u M^2} \Bigg] \Bigg\}~, 
\label{LCSRT1}
\eea
\bea
T_2(q^2) &=& \left( \frac{1}{2 (m_{B^*}^2 - m_{K^*}^2)} \right) \left(\frac{m_b}{f_B m_B^2} \right)
 \, \exp \left(\frac{m_B^2}{M^2}\right)
\int_\delta^1 \frac{du}{u}\, \exp \left( - \frac{m_b^2 + p^2 u \bar u - q^2 \bar u}{u
M^2}\right)  \nnb \\
&&\times \Bigg\{ f_{K^*} m_{K^*} \left[ g_\perp^{(v)} (u)- \frac{p^2}{2 M^2}
g_\perp^{(a)} (u)\right]q^2 + \frac{m_b f_{K^*}^\perp \phi_\perp (u)}{2 u}( m_b^2 -
q^2 - p^2 u^2)\;\nnb \\
&&+\; f_{K^*} m_{K^*} \Bigg[ \frac{ (m_b^2 -q^2- p^2 u^2)}{2 u}  \nnb \nnb \\
&& \times \Bigg( \Phi_\parallel (u) +
u g_\perp^{(v)} (u) + \frac{(m_b^2 -q^2-p^2 u^2) g_\perp^{(a)}(u)}{4 u M^2}
\Bigg) \Bigg] \Bigg\}
\label{LCSRT2}
\eea
and 
\bea
T_3(q^2) &=& \frac{1}{4} \left( \frac{m_b}{f_B m_B^2} \right) \exp \left(\frac{m_B^2}{M^2}\right)
\int_\delta^1 \frac{\d u}{u}\, 
\exp \ga - \frac{m_b^2 + p^2 u \bar u - q^2 \bar u}{uM^2}\dr
 \nnb \\
&&\times \Bigg\{ m_{K^*} f_{K^*} \Bigg[ \frac{g_\perp^{(a)}(u)}{4} + \frac{(m_b^2
-q^2-p^2 u^2)}{4 u M^2} g_\perp^{(a)} (u)\Bigg] \nnb \\
&& - \; 2 m_{K^*} f_{K*} \Bigg[
\frac{g_\perp^{(v)}(u)}{2}(2-u) - \frac{p^2 g_\perp^{(a)}(u)}{2 M^2}
\Bigg] \nnb \\
&& + \; 2 m_{K^*} f_{K*} \Bigg[ \frac{\Phi_\parallel (u)}{u M^2} \Bigg( \frac{m_b^2 - q^2 - p^2 u^2}{u}  \nnb \\
&& + \; q^2 - M^2 +
\frac{u M^2}{2}\Bigg) \Bigg] + m_b f_{K^*}^\perp \phi_\perp(u) \Bigg\}~.
\label{LCSRT3}
\eea
According to the light-cone sum rules derived in Ref. \cite{Aliev:1996hb}, the form factor $A_0$ is not independent but is given by 
\begin{equation}
A_0(q^2)= \left( \frac{m_B + m_{K^*}}{2 m_{K^*}} \right) A_1(q^2) + \left( \frac{q^2-m_B^2+m_{K^*}^2}{2m_{K^*}(m_B + m_{K^*})} \right) A_2(q^2) \;. 
\label{A0A1A2}
\end{equation}



The above form factors depend on parameters related to the $B$ meson, namely the Borel parameter $M_B$, the continuum threshold $s_0^B$, the quark mass $m_b $ and the $B$ meson decay constant $f_B$.  Here we follow Ref. \cite{Aliev:1996hb} and use the following set of parameter values : $M_B^2=8~\mbox{GeV}^2$, $s_0^B=36~\mbox{GeV}^2$ and $m_b=4.8~\mbox{GeV}$. We compute $f_B$ using the sum rule given in Ref. \cite{Ali:1993vd} in order to reduce the sensitivity of the form factors to the $b$-quark mass \cite{Ball:1997rj}. The lower integration limit depends on the continuum threshold: $\delta=(m_b^2-q^2)/(s^B_0-q^2)$. The function $\Phi_{\parallel}(u)$ is defined as 
 \begin{equation}
\Phi_\parallel(u) = \frac{1}{2}\,\left[ \bar u \int_0^u\!\!
\d v\,\frac{\phi_\parallel(v)}{\bar v} - u \int_u^1\!\!
\d v\,\frac{\phi_\parallel(v)}{v}\right]  
\label{eq:bigphi}
\end{equation}
and to leading twist-$2$ accuracy,  $g_\perp^{(v)}(u)$ and $g_\perp^{(a)}(u)$ are also given in terms of the longitudinal twist-$2$ DA \cite{Ball:1997rj} :
\begin{equation}
g_{\perp}^{(v)}(u) = \frac{1}{2}\,\left[  \int_0^u\!\!
\d v\,\frac{\phi_\parallel(v)}{\bar v} + \int_u^1\!\!
\d v\,\frac{\phi_\parallel(v)}{v}\right] 
\label{eq:gvtw2}
\end{equation}
and
\begin{equation}
g_{\perp}^{(a)}(u) = 2 \left[  \bar{u}\int_0^u\!\!
\d v\,\frac{\phi_\parallel(v)}{\bar v} + u \int_u^1\!\!
\d v\,\frac{\phi_\parallel(v)}{v}\right] \;. 
\label{eq:gatw2}
\end{equation}
Therefore the form factors depend additionally on the longitudinal and transverse twist-$2$ DAs $\phi_{\parallel,\perp}$ of the $K^*$ meson as well as its decay constants $f_{K^*}^{\perp}$ and $f_{K^*}$ which we shall discuss in the next section.


 
\section{AdS/QCD Distribution Amplitudes}

Traditionally, DAs are determined using QCD sum rules \cite{Ball:1996tb,Ball:1998fj,Ball:1998ff,Ball:2007zt}  which predict the moments of the DAs: 
\begin{equation}
\langle \xi_{\parallel, \perp}^n \rangle_\mu = \int \d z \; \xi^n \phi_{\parallel,\perp} (z, \mu)  
\end{equation}
where we have now made explicit the dependence of the DAs on the renormalization scale $\mu$. In the standard sum rules approach, only the first two non-vanishing moments are available  so that the sum rules DAs are reconstructed as truncated Gegenbauer polynomials:  
\begin{equation}
\phi_{\parallel,\perp}(z, \mu) = 6 z \bar z \left\{ 1 + \sum_{j=1}^{2}
a_j^{\parallel,\perp} (\mu) C_j^{3/2}(2z-1)\right\} 
\label{phiperp-SR}
\end{equation}
where $C_j^{3/2}$ are the Gegenbauer polynomials and the coefficients $a_j^{\parallel,\perp}(\mu)$ are related to the moments $\langle \xi_{\parallel,\perp}^n \rangle_\mu$ \cite{Choi:2007yu}. These moments and coefficients are determined at a low scale $\mu=1$ GeV and can then  be evolved perturbatively to higher scales \cite{Ball:2007zt}.  As $\mu \to \infty$, these coefficients vanish and the DAs take their asymptotic shapes. 

Alternatively, the DAs can be obtained using AdS/QCD \cite{Ahmady:2013cva}. The AdS/QCD DAs are related to the light-front wavefunction of the $K^*$ meson which can be obtained by solving the holographic light-front Schroedinger equation \cite{deTeramond:2008ht,Brodsky:2008kp} for mesons.  In Ref. \cite{Ahmady:2013cva}, we have shown that the twist-$2$ AdS/QCD DAs are given by
\begin{equation}
\phi_\parallel(z,\mu) =\frac{N_c}{\pi f_{K^*} m_{K^*}} \int \d
r \mu
J_1(\mu r) [m_{K^*}^2 z(1-z) + m_f m_{s} -\nabla_r^2] \frac{\phi_L(r,z)}{z(1-z)} \;,
\label{phiparallel-phiL}
\end{equation}
\begin{equation}
\phi_\perp(z,\mu) =\frac{N_c }{\pi f_{K*}^{\perp}} \int \d
r \mu
J_1(\mu r) [m_s - z(m_s-m_{\bar{q}})] \frac{\phi_T(r,z)}{z(1-z)} 
\label{phiperp-phiT}
\end{equation}
where $\phi_{\lambda=L,T}(r,z)$ are the AdS/QCD holographic light-front wavefunctions of the $K^*$ meson and $r$ the transverse distance between the quark and antiquark.  Explicitly, the holographic wavefunction is  given by \cite{Vega:2009zb}
\begin{equation}
\phi_{\lambda} (z,\zeta)= \mathcal{N}_{\lambda}
\frac{\kappa}{\sqrt{\pi}}\sqrt{z(1-z)} \exp
\left(-\frac{\kappa^2 \zeta^2}{2}\right)
\exp\left \{-\left[\frac{m_s^2-z(m_s^2-m^2_{\bar{q}})}{2\kappa^2 z (1-z)} \right]
\right \} \label{AdS-QCD-wfn}
\end{equation}
where $\zeta=\sqrt{z(1-z)} r$ is the light-front variable  that maps onto the fifth dimension of AdS space \cite{deTeramond:2008ht}. The AdS/QCD wavefunction given by  Eq. \eqref{AdS-QCD-wfn} is obtained using a quadratic \cite{Brodsky:2013ar,Brodsky:2013npa}  dilaton in AdS in order to simulate confinement in physical spacetime.  In that case, the parameter $\kappa=m_{K^*}/\sqrt{2}$ . As discussed in reference \cite{Forshaw:2012im}, the normalization $\mathcal{N}_{\lambda}$ of the AdS/QCD wavefunction is fixed according to the polarization of the meson. 

\begin{figure}
\centering
{\includegraphics[width=.80\textwidth]{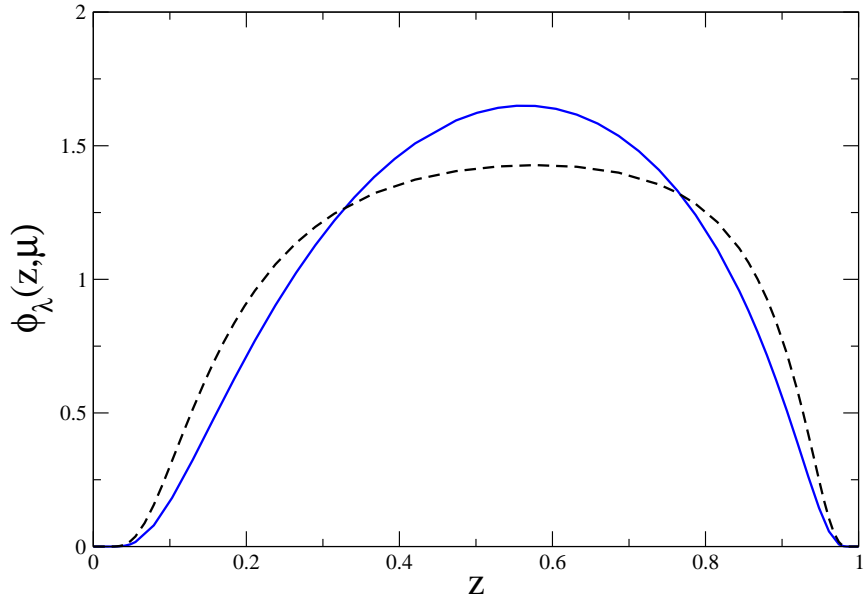} }
\caption{The AdS/QCD twist-$2$ DAs. Solid blue: longitudinal ($\lambda=\parallel$) twist-$2$ DA. Dashed black: transverse ($\lambda=\perp$) twist-$2$ DA. } \label{fig:DAs}
\end{figure}

Note that both DAs are normalized, i.e.
\begin{equation}
\int_0^1 \d z \; \phi_{\perp,\parallel} (z, \mu)=1 
\end{equation}
so  that the decay constants are given by \cite{Ahmady:2013cva}
\begin{equation}
f_{K^*}  = \frac{N_c}{m_{K^*}\pi}  \int_0^1 \d z
\left.[z(1-z)m^{2}_{K^*} + m_{\bar{q}} m_{s} -\nabla_{r}^{2}]
\frac{\phi_L(r,z)}{z(1-z)}
\right|_{r=0} 
\label{vector-decay}
\end{equation}
and
\begin{equation}
f_{K^*}^{\perp}(\mu) =\frac{N_c}{\pi} \int_0^1 \d z (m_{s} -z(m_s-m_{\bar{q}})) \int \mu J_1(\mu r)  \frac{\phi_T(r,z)}{z(1-z)}
\label{tensor-decay-mu} \;.
\end{equation}
We are thus able to make AdS/QCD predictions for the decay constants using Eqns. \eqref{vector-decay} and \eqref{tensor-decay-mu}. Using constituent quark masses, i.e. $m_{\bar{q}}=0.35$ GeV and $m_{s}=0.48$ GeV, we obtain $f_{K^*}=225$ MeV and $f_{K^*}^{\perp} (\mu)=119$ MeV for $\mu \ge 1$ GeV. We point out that we choose constituent quark masses since they lead to a prediction for the ratio $f_{K^*}^{\perp}/f_{K^*}$ that is closest to the corresponding sum rules and lattice predictions \cite{Ahmady:2013cva}. Note also that our AdS/QCD DAs, shown in Fig. \ref{fig:DAs}, also hardly depend on $\mu$ for $\mu \ge 1$ GeV.  

 
\section{Results}
\begin{figure}
\centering
{\includegraphics[width=.80\textwidth]{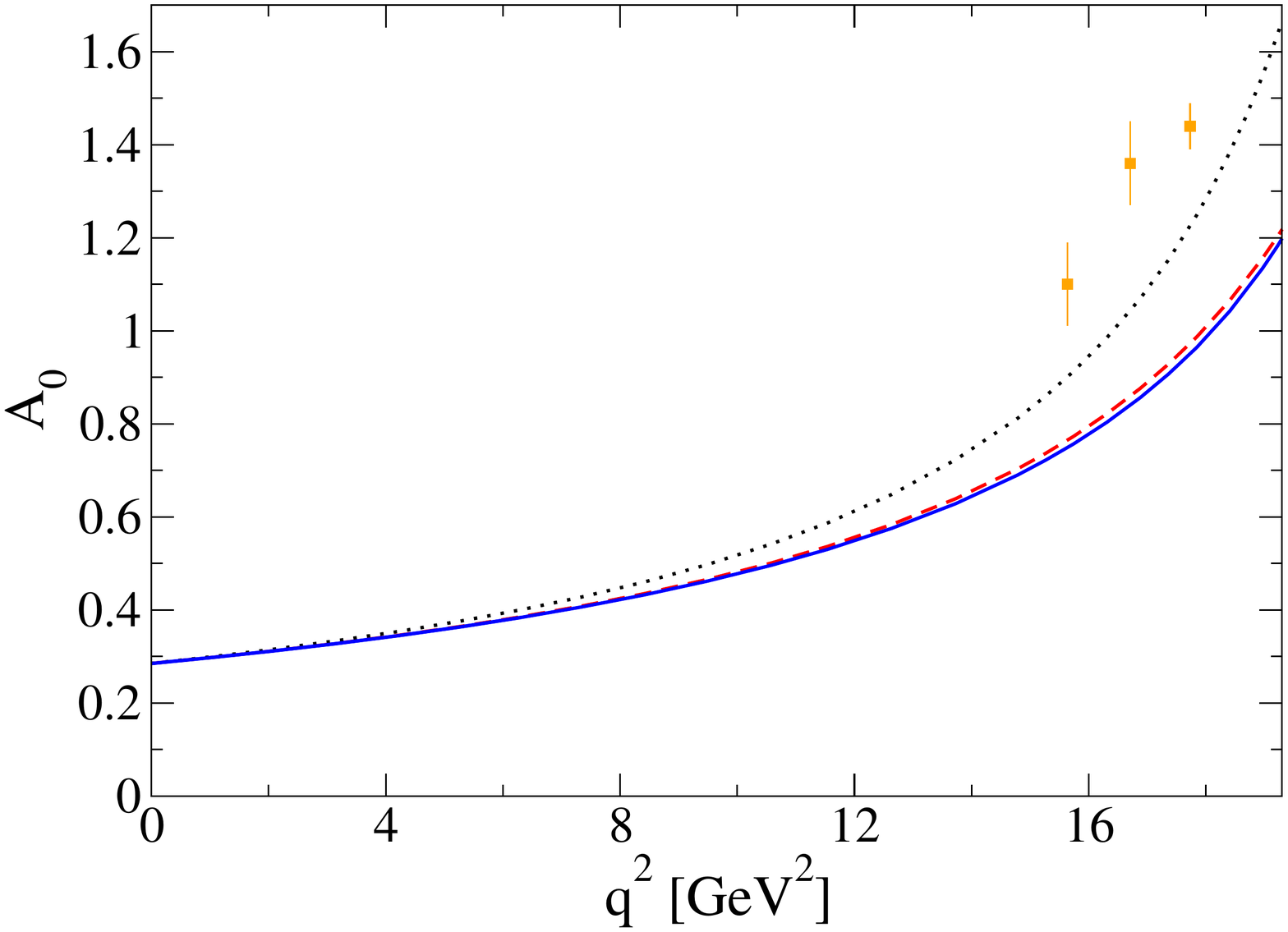} }
\caption{The axial-vector form factor $A_0$. Solid blue: AdS/QCD. Dashed red: AdS/QCD Fit.  Dotted black: Fit to AdS/QCD and lattice. Orange data points: lattice data.} \label{fig:A0}
\end{figure}

\begin{figure}
\centering
{\includegraphics[width=.80\textwidth]{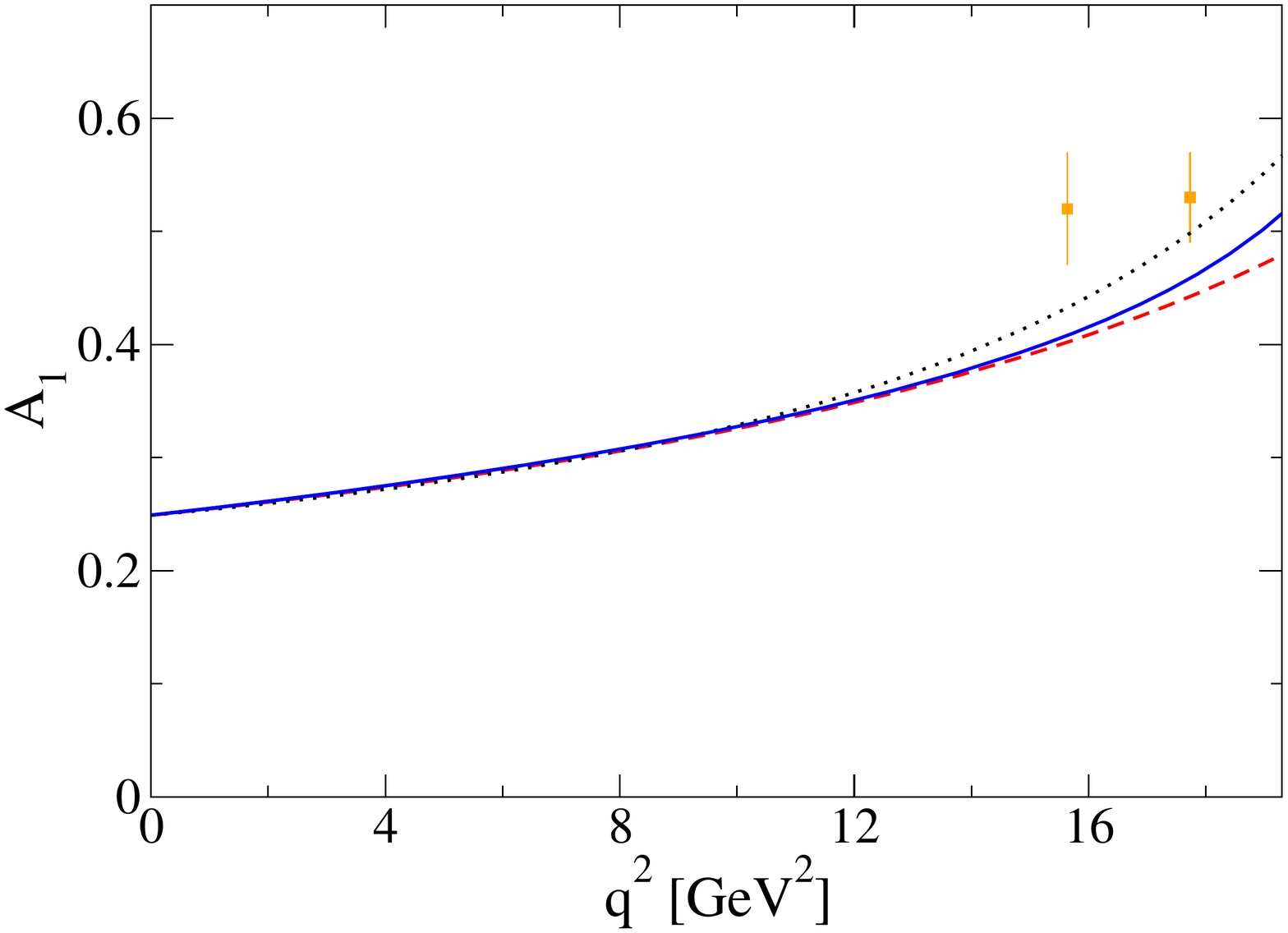} }
\caption{The axial-vector form factor $A_1$. Solid blue: AdS/QCD. Dashed red: AdS/QCD Fit. Dotted black: Fit to AdS/QCD and lattice. Orange data points: lattice data.} \label{fig:A1}
\end{figure}

\begin{figure}
\centering
{\includegraphics[width=.80\textwidth]{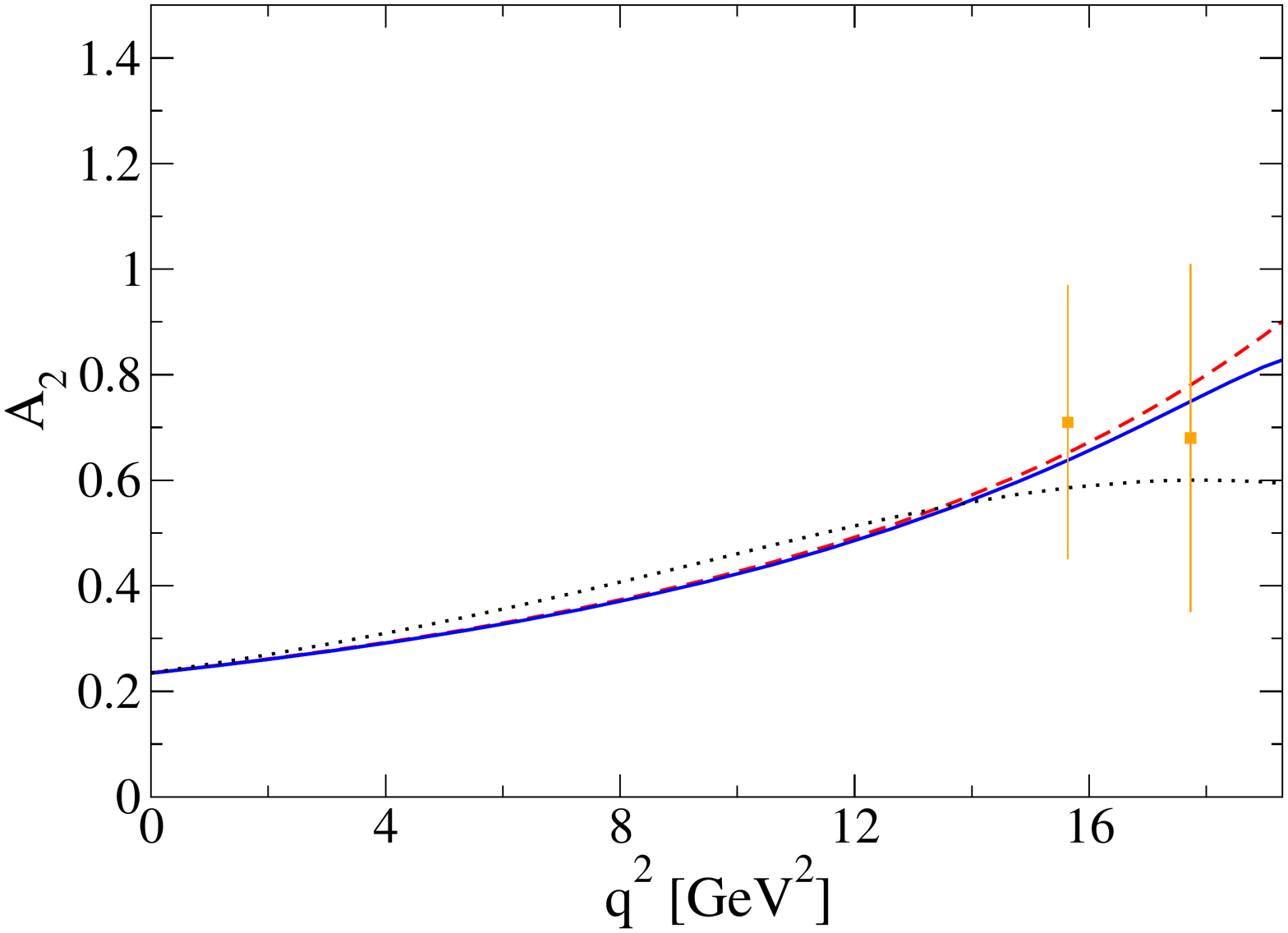} }
\caption{The axial-vector form factor $A_2$. Solid blue: AdS/QCD. Dashed red: AdS/QCD Fit. Dotted black: Fit to AdS/QCD and lattice. Orange data points: lattice data.} \label{fig:A2}
\end{figure}

\begin{figure}
\centering
{\includegraphics[width=.80\textwidth]{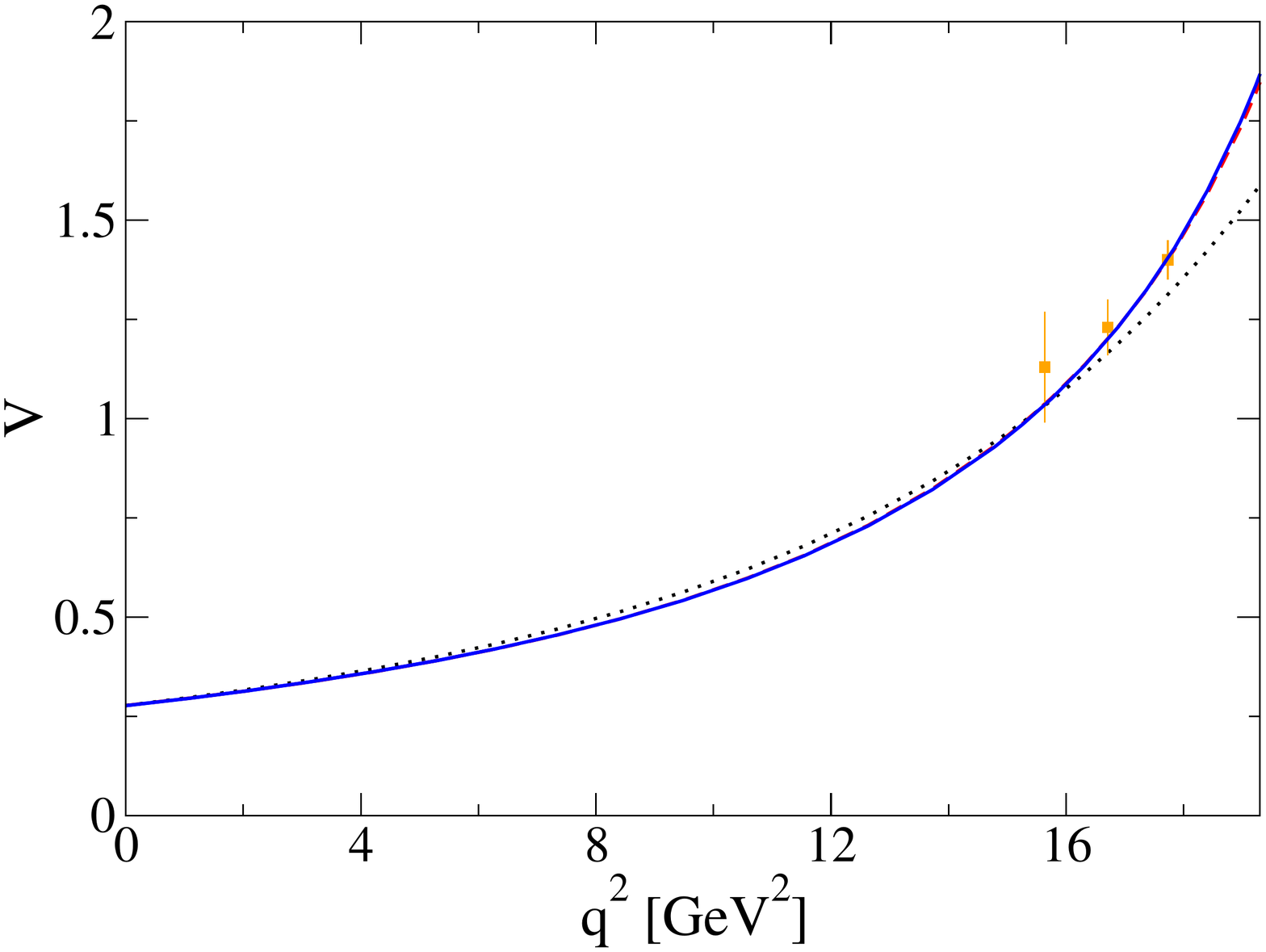} }
\caption{The vector form factor $V$. Solid blue: AdS/QCD. Dashed red: AdS/QCD Fit. Dotted black: Fit to AdS/QCD and lattice. Orange data points: lattice data. } \label{fig:V}
\end{figure}

\begin{figure}
\centering
{\includegraphics[width=.80\textwidth]{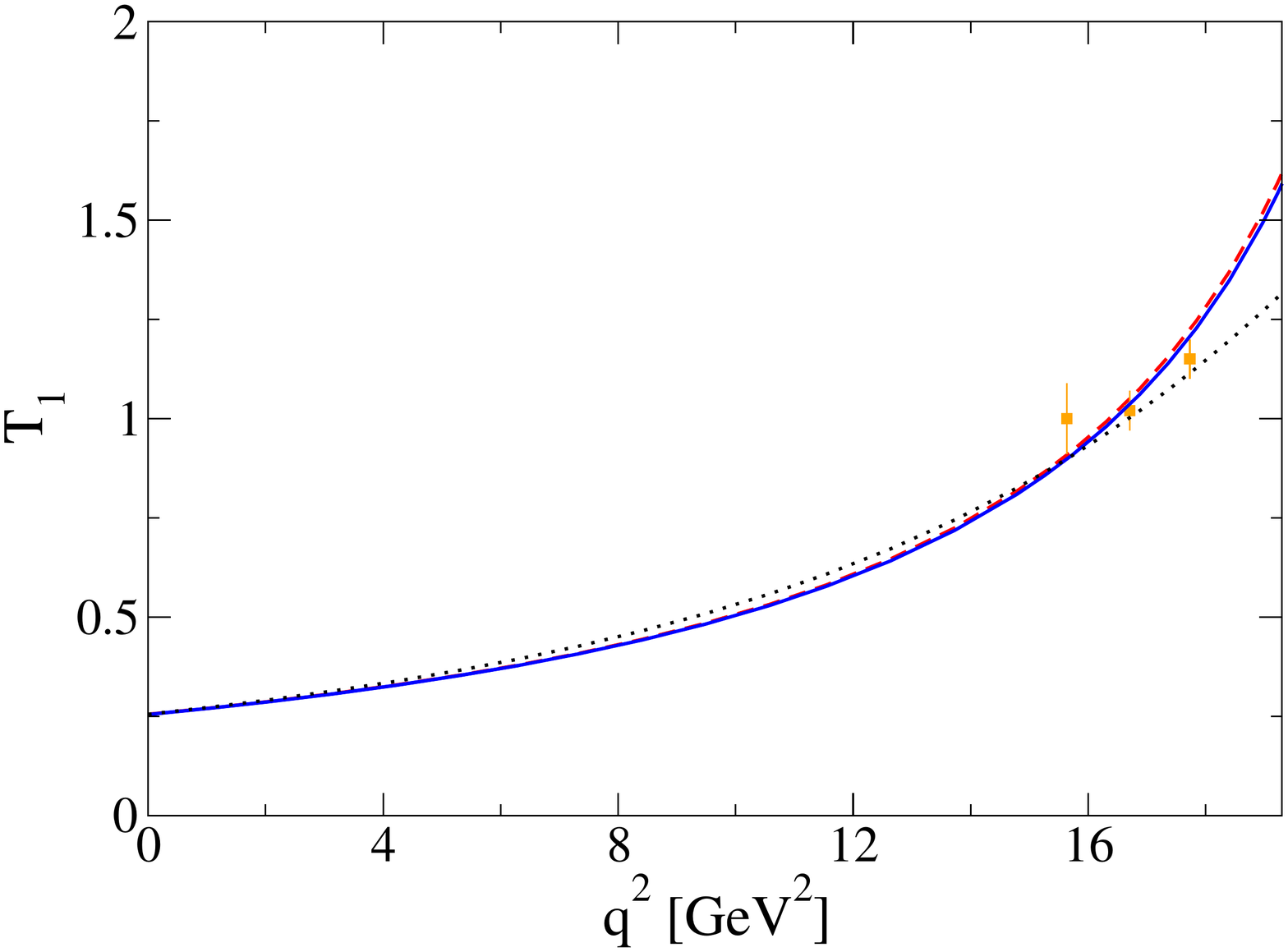} }
\caption{The tensor form factor $T_1$. Solid blue: AdS/QCD. Dashed red: AdS/QCD Fit. Dotted black: Fit to AdS/QCD and lattice. Orange data points: lattice data.} \label{fig:T1}
\end{figure}

\begin{figure}
\centering
{\includegraphics[width=.80\textwidth]{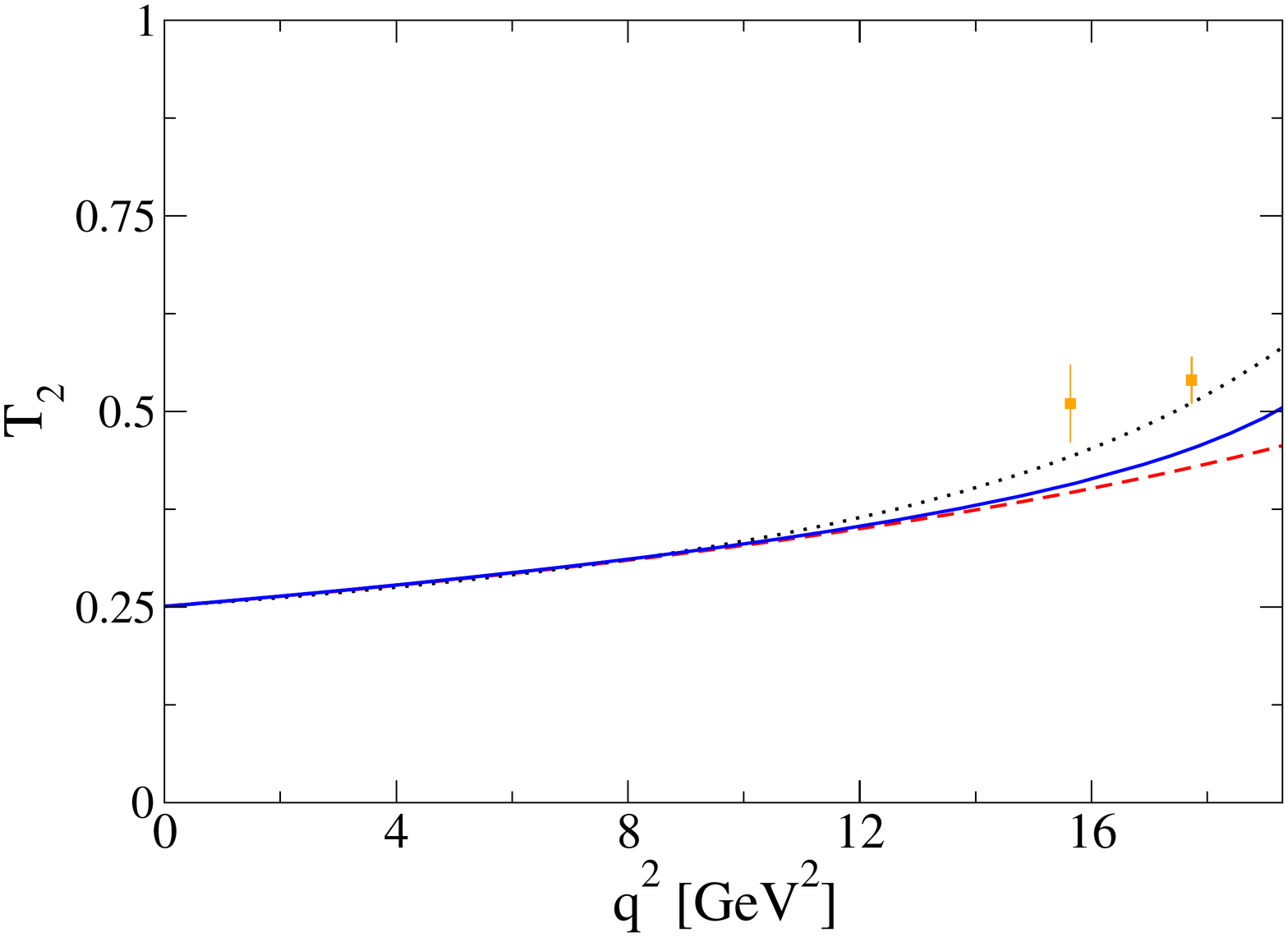} }
\caption{The tensor form factor $T_2$. Solid blue: AdS/QCD. Dashed red: AdS/QCD Fit. Dotted black: Fit to AdS/QCD and lattice. Orange data points: lattice data.} \label{fig:T2}
\end{figure}

\begin{figure}
\centering
{\includegraphics[width=.80\textwidth]{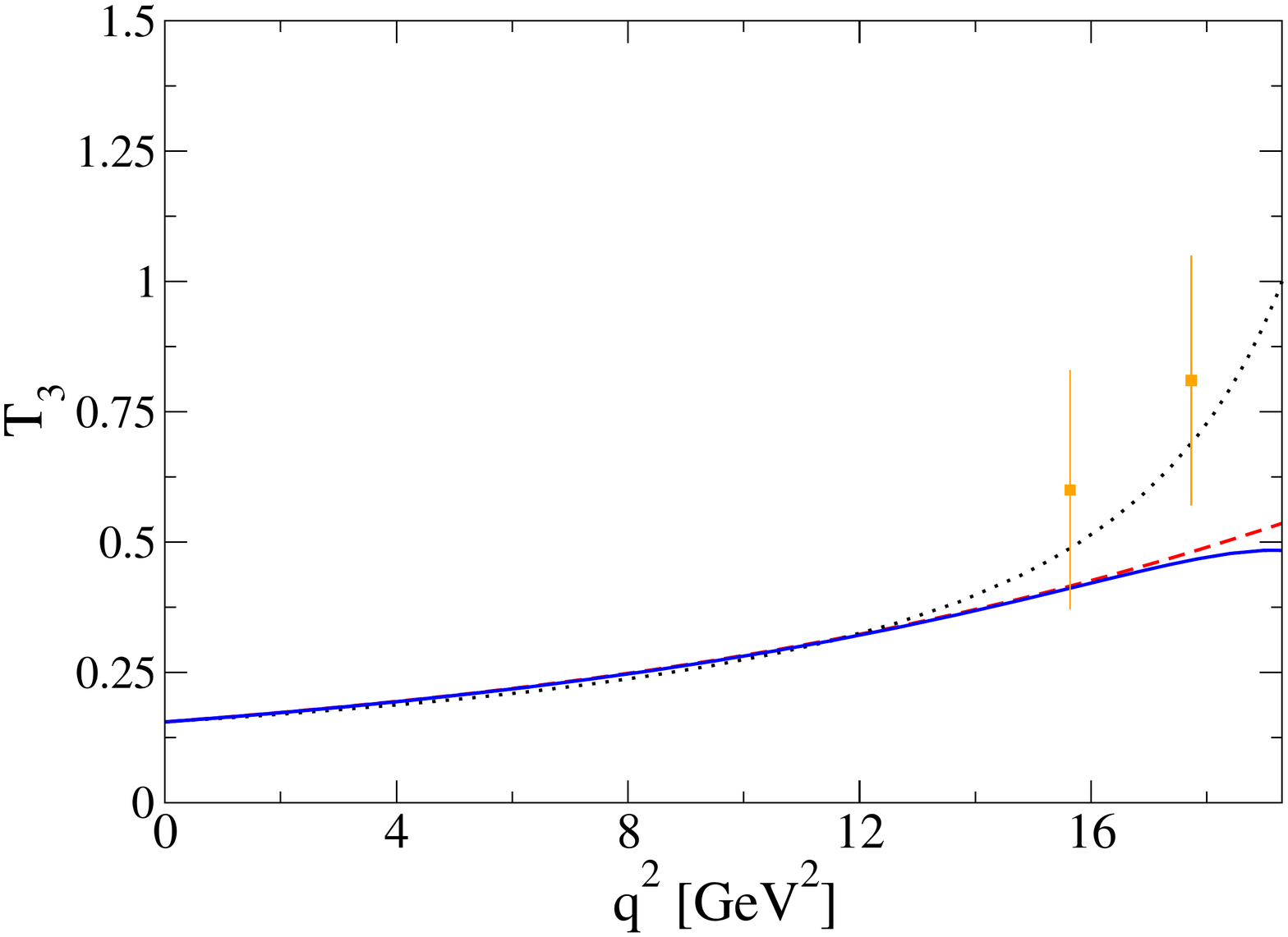} }
\caption{The tensor form factor $T_3$. Solid blue: AdS/QCD. Dashed red: AdS/QCD Fit. Dotted black: Fit to AdS/QCD and lattice. Orange data points: lattice data.} \label{fig:T3}
\end{figure}

Having specified the DAs and decay constants, we are now in a position to compute the form factors using the light-cone sum rules given by Eqns. \eqref{LCSRV} to \eqref{LCSRT3}. In Figs. \ref{fig:A0} to \ref{fig:T3}, we show our predictions for the seven form factors. In each figure, the solid blue curve is generated using our AdS/QCD DAs and decay constants.  Restricting the AdS/QCD predictions to low-to-intermediate $q^2$ (in practice, we take $0 \le q^2 \le 16~\mbox{GeV}^2$) for each form factor, we fit the parametric form 
\begin{equation}
F(q^2)=\frac{F(0)}{1- a (q^2/m_B^2) + b (q^4/m_B^4)}
\label{FitFF}
\end{equation}
to our predictions. The fitted values of the parameters $a$ and $b$ are given in Table \ref{tab:abAdS} and the resulting form factors are shown as the dashed red curves in Figs. \ref{fig:A0} to \ref{fig:T3}. 
We repeat the fits by including  the most recent unquenched lattice data  of Ref. \cite{Horgan:2013hoa}.  We use the data set obtained using the smallest lattice spacing. The fitted values of $a$ and $b$ are collected in Table \ref{tab:abAdSlat} and the resulting form factors are shown as the dotted black curves in Figs. \ref{fig:A0} to \ref{fig:T3}.

\begin{table}[h]
\begin{center}
\[
\begin{array}
[c]{|c|c|c|c|}\hline
F&F(0)&a&b\\ \hline
A_0&0.285& 1.158 & 0.096 \\ \hline
A_1&0.249& 0.625 &-0.119  \\ \hline
A_2&0.235&  1.438&0.554 \\ \hline
V&0.277& 1.642& 0.600\\ \hline
T_1&0.255& 1.557& 0.499\\ \hline
T_2 & 0.251 &0.665  &-0.028  \\ \hline
T_3& 0.155 & 1.503&0.695 \\ \hline
\end{array}
\]
\end{center}
\caption {The values of the form factors at $q^2=0$ together with the fitted parameters $a$ and $b$. The values of $a$ and $b$ are obtained by fitting Eq. \eqref{FitFF} to the AdS/QCD predictions for low-to-intermediate $q^2$. }
\label{tab:abAdS}
\end{table}

\begin{table}[h]
\begin{center}
\[
\begin{array}
[c]{|c|c|c|c|}\hline
F&F(0)&a&b\\ \hline
A_0&0.285& 1.314&0.160 \\ \hline
A_1&0.249&  0.537&-0.403 \\ \hline
A_2&0.235& 1.895&1.453 \\ \hline
V&0.277& 1.783&0.840\\ \hline
T_1&0.255&1.750 &0.842 \\ \hline
T_2 & 0.251 &0.555 &-0.379\\ \hline
T_3& 0.155 &1.208 &-0.030 \\ \hline
\end{array}
\]
\end{center}
\caption {The values of the form factors at $q^2=0$ together with the fitted parameters $a$ and $b$. The values of $a$ and $b$ are obtained by fitting Eq. \eqref{FitFF} to both the AdS/QCD predictions for low-to-intermediate $q^2$ and the lattice data at high $q^2$. }
\label{tab:abAdSlat}
\end{table}

Finally we compute the differential branching fraction\footnote{We have summed over the two possible final lepton polarizations.} given by \cite{Aliev:1996hb}


\begin{eqnarray}\label{gammaK}\nonumber
&&\frac{d\mathcal{B}}{dq^{2}}=\tau_{B}\frac{G_{F}^{2}\alpha^{2}}{2^{11}\pi^{5}}\frac{|{V_{tb}V_{ts}^{*}|^{2}\sqrt{\lambda}v}}{3m_{B}}((2m_{\mu}^{2}+m_{B}^{2}s)[16(|A|^{2}+|C|^{2})m_{B}^{4}\lambda+2(|B_{1}|^{2}+|D_{1}|^{2})\\ \nonumber
&&\times\frac{\lambda+12rs}{rs}+2(|B_{2}|^{2}+|D_{2}|^{2})\frac{m_{B}^{4}\lambda^{2}}{rs}-4[\Re \mbox{e}(B_{1}B_{2}^{*})+\Re\mbox{e}(D_{1}D_{2}^{*})]\frac{m_{B}^{2}\lambda}{rs}(1-r-s)]\\ \nonumber
&&+6m_{\mu}^{2}[-16|C|^{2}m_{B}^{4}\lambda+4\Re \mbox{e}(D_{1}D_{3}^{*})\frac{m_B^2\lambda}{r}-4\Re \mbox{e}(D_{2}D_{3}^{*})\frac{m_{B}^{4}(1-r)\lambda}{r}+2|D_{3}|^{2}\frac{m_{B}^{4}s\lambda}{r}\\ \nonumber
&&-4\Re \mbox{e}(D_{1}D_{2}^{*})\frac{m_{B}^{2}\lambda}{r}-24|D_{1}|^{2}+2|D_{2}|^{2}\frac{m_{B}^{4}\lambda}{r}(2+2r-s)])
\end{eqnarray}
where $\lambda=1+r^{2}+s^{2}-2r-2s-2rs$, with $r=m_{K}^{2}/m_{B}^{2}$ and $s=q^{2}/m_{B}^{2}$. The final muon has  mass $m_{\mu}$ and velocity $v=\sqrt{1-4 m_{\mu}^{2}/q^{2}}$. We take $\tau_B$ as the average of the lifetimes of the $B^{\circ}$ and $B^+$. The differential branching fraction depends on the following combinations of form factors: 

\begin{eqnarray}\label{A}
A=C_{9}^{eff}\left(\frac{V}{m_{B}+m_{K^*}}\right)+4C_{7}\frac{m_{b}}{q^{2}}T_{1} \;,
\end{eqnarray}

\begin{eqnarray}\label{B1}
B_{1}=C_{9}^{eff}(m_{B}+m_{K^*})A_{1}+4C_{7}\frac{m_{b}}{q^{2}}(m_{B}^{2}-m_{K^*}^{2})T_{2} \;,
\end{eqnarray}

\begin{eqnarray}\label{B2}
B_{2}=C_{9}^{eff}\left(\frac{A_{2}}{m_{B}+m_{K^*}}\right) + 4C_{7}\frac{m_{b}}{q^{2}}\left(T_{2}+\frac{q^{2}}{m_{B}^{2}-m_{K^*}^{2}}T_{3}\right) \;,
\end{eqnarray}

\begin{eqnarray}\label{C}
C=C_{10}\left(\frac{V}{m_{B}+m_{K^*}} \right)\;,
\end{eqnarray}

\begin{eqnarray}\label{D1}
D_{1}=C_{10}(m_{B}+m_{K^*})A_{1} \;,
\end{eqnarray}

\begin{eqnarray}\label{D2}
D_{2}=C_{10}\left(\frac{A_{2}}{m_{B}+m_{K^*}} \right)
\end{eqnarray}
and
\begin{eqnarray}\label{D3}
D_{3}=-C_{10}\frac{2m_{K^*}}{q^{2}}\left( \left(\frac{m_B + m_{K^*}}{2 m_{K^*}} \right) A_1 - \left( \frac{m_B - m_{K^*}}{2 m_{K^*}}\right) A_2-A_{0}\right)
\end{eqnarray}
where $C_9^{eff}=C_9 + Y(q^2)$ with $C_9$, $C_7$ and $C_{10}$ being the Standard Model Wilson coefficients given in  Ref. \cite{Altmannshofer:2008dz} . The function $Y(q^2)$ is also given explicitly in Ref.  \cite{Altmannshofer:2008dz}.

\begin{figure}
\centering
{\includegraphics[width=1.0\textwidth]{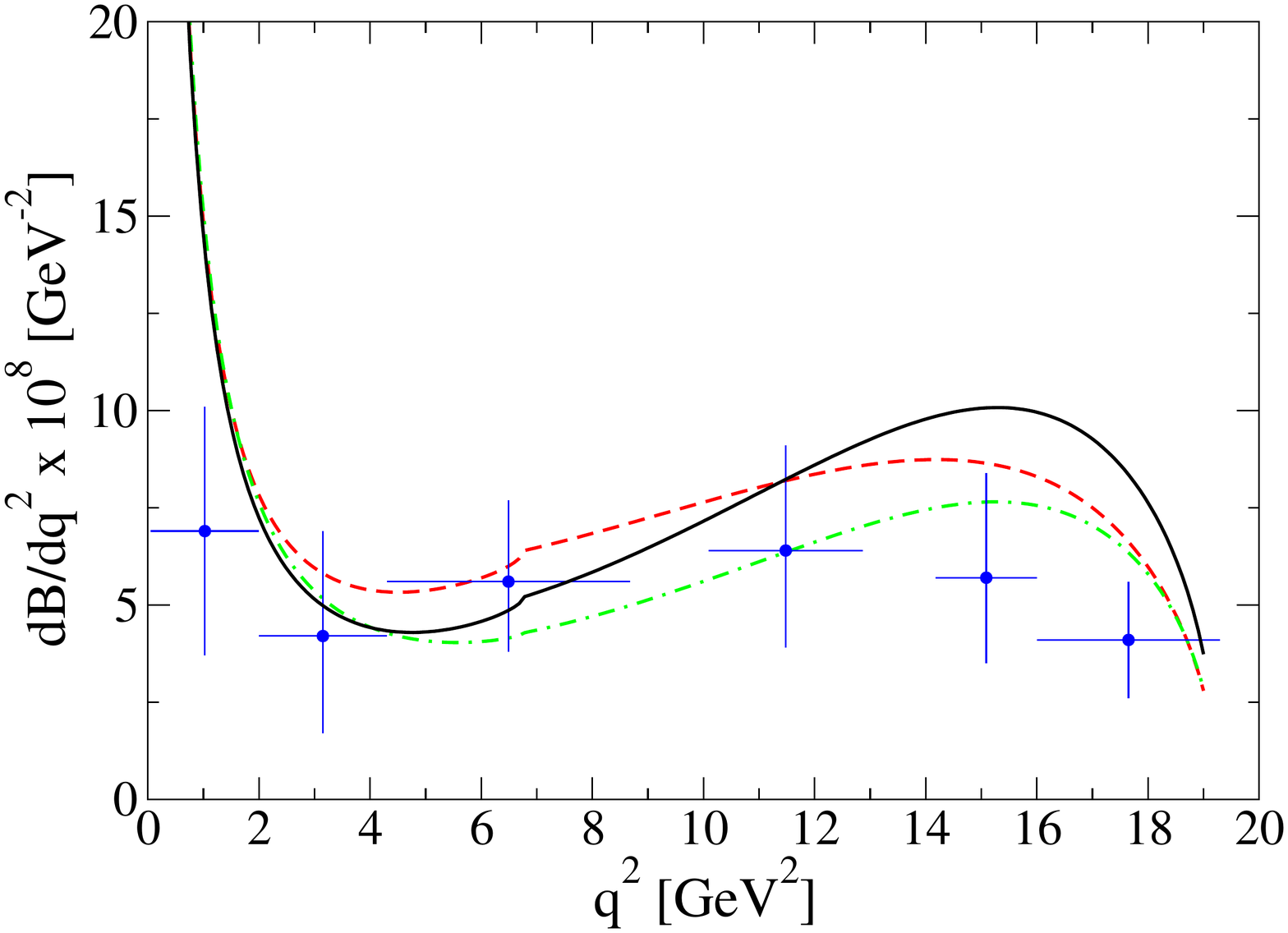} }
\caption{The differential branching fraction as a function of $q^2$. For the LHCb data, we average over the data for $B^+ \to K^{*+} \mu^+ \mu^-$ given in   Ref. \cite{Aaij:2012cq} and the data for $B^\circ \to K^{*\circ} \mu^+ \mu^-$ given in Ref. \cite{LHCb-CONF-2012-008}. Dashed red : AdS/QCD fit. Solid black: AdS/QCD $+$ lattice fit. Dot-dashed green: AdS/QCD $+$ lattice $+$ NP in the Wilson coefficient $C_9$.} \label{fig:decaywidth}
\end{figure}

In Fig. \ref{fig:decaywidth}, we show our predictions for the differential branching ratio. The dashed red curve is generated by using the form factors that are fitted to only the AdS/QCD predictions. As can be seen, the curve somewhat overshoots the LHCb data at  high $q^2$ especially in the $14~\mbox{GeV}^2 \le q^2 \le 16~\mbox{GeV}^2$ bin. Using the form factors that fit both the AdS/QCD predictions and the lattice data, we obtain the solid black curve. Rather surprisingly, adding the lattice data in our form factor fits worsens the disagreement at high $q^2$. However, this disagreement at high $q^2$ is consistent with the findings of Ref. \cite{Horgan:2013pva} in the authors studied the possibility of new physics in the Wilson coefficients $C_9$ and $C_9^{\prime}$. Indeed, by adding a negative new physics contribution, i.e. taking $C_9^{\mbox{\tiny{NP}}}=-1.5$ \cite{Descotes-Genon:2013wba} (we keep $C_9^\prime=0$ as in the SM), we find a better fit to the data at high $q^2$as shown by the dot-dashed green curve of Fig. \ref{fig:decaywidth}. 


 By integrating Eqn. \eqref{gammaK} over $q^2$ and excluding the regions of the narrow charmonium resonances, we obtain a total branching fraction of $1.56 \times 10^{-6}(1.55 \times 10^{-6})$ when including (excluding) the lattice data compared to the LHCb measurement $(1.16 \pm 0.19)\times 10^{-6}$. In both cases, we overestimate the total branching fraction. With a new 
physics contribution to $C_9$, we obtain a total branching fraction of $1.35 \times 10^{-6}$ in agreement with the LHCb data.

\section{Conclusions}
We have computed the seven $B \to K^*$ transition form factors using QCD light-cone sum rules with AdS/QCD DAs. We have provided two sets of parametrizations for the form factors: a first set that fits our AdS/QCD predictions at intermediate-to-high recoil and a second set that fits both our AdS/QCD predictions as well as the most recent lattice data at low recoil. The first set gives a good description of the data except at high $q^2$ where our prediction overshoots the data.  The second set worsens the disagreement at high $q^2$. We looked into the possibility of a new physics contribution in the Wilson coefficient $C_9$ in order to improve agreement at high $q^2$.

\section{Acknowledgements}
This research is supported by the Natural Sciences and Engineering Research Council of Canada (NSERC). We thank T. Aliev and M. Wingate for useful correspondence.

\bibliographystyle{apsrev}
\bibliography{AdSFFKstar}

\end{document}